# "Teaching" information ethics, or co-constructing shared values and practices?

Lessons learned from a curriculum on Scientific Integrity, Research Ethics and Information Ethics in STEM disciplines.

Thomas Baudel

IBM France Lab, Université Paris-Saclay. baudelth@fr.ibm.com

We present the motivation, design, outline and lessons learned from an online course in scientific integrity, research ethics and information ethics provided to over 2000 doctoral and engineering students in STEM fields, first at the University Paris-Saclay, and now expanded to an online MOOC available to students across the world, in English. Unlike a course in scientific domains, meant to provide students with methods, tools and concepts they can apply in their future career, the goal of such a training is not so much to equip them, but to make them aware of the impact of their work on society, care about the responsibilities that befall on them, and make them realize not all share the same opinions on how should technology imprint society. While we provide conceptual tools, this is more to sustain interest and engage students. We want them to debate on concrete ethical issues and realize the difficulty of reconciling positions on contemporary dilemma such as dematerialized intellectual property, freedom of expression online and its counterparts, the protection of our digital selves, the management of algorithmic decision, the control of autonomous systems, and the resolution of the digital divide. As a bold shortcut, our course is about introducing and motivating Hegelian dialectics in STEM curricula, usually more bent on an Aristotelian perspective.

## 1 INTRODUCTION

### 1.1 A concrete need for a curriculum in Scientific integrity and Research Ethics for information technologists

In May 2016, a ministerial decree mandated doctoral schools in France to provide training in scientific integrity and research ethics to their PhD Students. This decree did not come as a surprise, as most countries around the world make similar requirements, think tanks and institutions provide reports and curriculum suggestions, and both the media and political institutions show increasing preoccupation about the role of science and technology in shaping the future of our societies. What may be more surprising is that a curriculum proposition for the doctoral college in Information Sciences and Technologies at University Paris-Saclay came from an external member of the doctoral school council, working in the industry. As a researcher in the industry, plenty of the concerns that justify such a training have emerged from the design of new products our customers demand. In trying to answer their needs or open new opportunities, we have stumbled upon new types of issues, that need to be addressed in ways much different than our scientific training had given us. Designing products that we put in the hands of our customers and that affect millions of users forbids us to look sideways at uneasy questions, if only for liability reasons if we assume purely cynical motivations. We hire new recruits and need to have them ready to face those questions, which makes this training all the more valuable. Together with a think-tank from the university Paris-Saclay, led by Christine Froidevaux, we worked on the setup of an online course that was first provided to the doctoral students in computer science at University Paris-Saclay in February 2018, and has been extended as an online MOOC available to everyone at the onset of 2021.

A separate motivation for creating such a course is the recent wave of interest around AI Ethics. From an epistemological standpoint, there are no conceptual reasons to isolate such a subdomain of information ethics: it makes more sense to organize the domain by the human areas that are affected by technology than by the properties of some particular information devices, independently of the context in which they are used. However, the buzzwords have attracted a significant following, and raised a sustained interest in the general public. Riding this wave creates additional opportunities to entice our public in, first, realizing the depth of the responsibilities their work on furthering information technologies entails, and second, acquiring conceptual tools and the work practices needed to assume them.

### 1.2 A fundamental need to rethink our ethical frameworks for the information age

Before entering the pragmatic considerations that led us to the design of this course, we may first ponder over the reason information ethics has recently risen to a major preoccupation, not just in the Academia, but also the general public. Dozens, perhaps even hundreds, of initiatives, think tanks, normalization and regulations projects regarding information technologies have sprouted over the past decade, to the point there is no use in listing them. One of the most notable aspects of this movement is that many, if not most, of these initiatives originate from industry actors rather than from regulatory bodies or more traditional actors of the civil society involved in setting social norms and regulations. While we should not discard the occurrence of some "Ethics washing", we believe the need arises from the concrete observations that technology creators are facing more and more often a new class of problems that affect their work area, and which are not properly addressed by classical engineering and scientific methods. Those problems involve consensus building, communication, and a dialectical debate, building a sort of "proof by consensus" that a classically trained scientist, having been ingrained with opposing επιστήμη to δόξα, should abhor. And yet, it is no longer possible to ignore the impact their work has on the very fabric of society, and thus the need for educating and consensus building. Two main factors explain this relatively new, but massive, phenomenon:

One is very visible: the diffusion of information technologies in every aspect of our lives. Part of what makes our humanity is transitioning from a world made of atoms to a world made of bits, which Luciano Floridi describes as the *infosphere* [7]. The impact of this transition, at varying speeds across society and domains, forces the technologists who make it happen to ponder over its consequences and directions. The ethical implications of this transition impose a fundamental shift in research ethics: whereas *addressing a dilemma in bioethics is akin to answering the question "who are we?" ; in technology, in particular information technologies, issues tackle more the interrogation: "where are we going?".*

The second factor pertains to the difference between natural objects and informational objects. An interesting argument by Rafael Capurro explains well its implications [2]: If we follow Martin Heidegger in considering that the fundamental characteristic of our existence is that of "being there", that is to say, a conscious entity bounded in time and space, then we have to also consider that an informational agent (some of which may be qualified as "AI") is not bound by the same constraints: it is not specifically located in space, can be perfectly cloned without the possibility of distinguishing the clone from the original, and is not subject to the second law of thermodynamics that would impose an implicit expiry date. As a consequence, we cannot say that it "is there" in the same sense that we "are there". Hence, if our ethical systems and practices are to be determined by the conditions of our existence, the shift to a digital world implies that the very foundations of our ethical frameworks are to be rethought in this new situation. *The digital revolution therefore constitutes not just a host of new ethical challenges, but perhaps a fundamental shift in our approaches on ethical thinking.*

In this article, we first synthesize the various viewpoints and motivations for providing this training that we have considered to refine the pedagogical goals. Next, we present the course outline and its mode of operation, before summarizing the lessons learned, in the form of statistics and general feedback. In conclusion, we present how, based on our experience, should AI Ethics concerns be integrated in this type of wide-scale curriculum, and the future directions we aim at providing for this course.

## 2 CONSTRAINTS AND GOALS OF THE COURSE

### 2.1 Training methods in Research Ethics, Compliance, and Information Ethics

#### 2.1.1 Scientific integrity and research ethics training

The course mandate originates from the Corvol report [3], a report to the French Ministry of Research to propose policies to enforce scientific integrity in the French research ecosystem. This report is largely informed by the current practice of research institutions worldwide. Those mandate similar training, and we refer to the report itself for an extended bibliography. Among its conclusions, the report proposes a mandatory training to research personnel, including doctoral students, sensitizing them to the issues of scientific integrity and research ethics, with simple case-based courses evaluated with quiz-like questionnaires.

Our course follows roughly this form, but it is more demanding than what the Corvol report suggests, because of more specific ambitions we wanted to introduce and that are presented later. To adapt the Corvol mandate to the specific needs of Computer and Information sciences, the CERNA-Allistene, a think tank focused on information ethics in the Academia, proposed the outline of a course that has greatly inspired our course [4]. There is indeed a need to address an ethics of technology differently than what bioethics, the foundational discipline of research ethics, as pointed earlier and by Floridi [8].

#### 2.1.2 Compliance training in highly regulated industries

Highly regulated industry, such as finance, insurance, health… are required to maintain and enforce proper deontological and ethical standards in their workforce. Not only must these industries provide training, they must also evaluate the effectiveness of this training, and maintain minimal standards on several indicators. These requirements have led to the establishment of a strong industry practice, under the name of "Ethics and Compliance". Cross-industry best practices are largely shared by longstanding institutions, such as the Ethics and Compliance Initiative [6].

For compliance training, the preferred format involves, first a short booklet of general ethical guidelines that all employees must read, and, second, most interestingly, a yearly 1-hour mandatory online training session. This session presents a few scenarios exhibiting a deontological or ethical uncertainty, sufficiently generalized to be applicable to everyone in the company: a manager asks to share a password, a customer asks for a feature that would let them work around legal obligations… The case is like a "book where you are the hero" and the test-taker is invited, with quizzes, to present how they would react in those difficult circumstances. Then, numerous resources in relation to the case are provided to address similar situations they might encounter. This technique appears very effective, can reach the full enterprise, and has proven results in maintaining good ethical standards. The scenarios are renewed every year to address different types of ethical and deontological issues.

Parts of our curriculum is inspired by this approach, and we have attempted to reproduce some of its positive aspects, by presenting case studies and scenarios. In the long term, ethical and deontological training in research and engineering should most likely take the same shape. However, it appears that:
- We need a comprehensive coverage of a large ethical and deontological landscape and create a mindset on ethical thinking. This undertaking takes much more than light and entertaining courses taken every year.
- Our target population is much more focused on abstraction and expects some intellectually challenging content to sustain interest.
- Providing a multitude of scenarios is a daunting task, it requires a major investment that is perhaps too early to make.

### 2.1.3 Curricula on Information Ethics, Security, Privacy & AI Ethics

In light of the many initiatives on AI Ethics, the public debate on the place of algorithms in society, the role of social networks, the future of work in an automated world…, we can now consider that a full Academic domain has emerged over the past decade that is concerned with the transformative impact of Information Sciences and Technologies on humans, society and nature [10]. Large internet companies are now hiring "privacy engineers", ethicists, sociologists to guide them. Requirements for these positions involve a blend of technical and humanities competences that is difficult to find, explaining the blooming of a training offer in Information Ethics, some [5] even attracting their audience with promises of 25% salary increase! Because the changes digital technologies will bring affect all of us, other initiatives involve providing free courses on AI technologies to the general public [9].

This relatively new offer questions our objectives: should we want to "compete" in this space, training specialists in AI Ethics and various sociological aspects of computing? Considering our course will be made mandatory for all students of a faculty, this would be inappropriate. Also, the blend of hype and marketing around ill-defined concepts that abounds in the area of AI Ethics, in our opinion, obscures the true challenges that our future generations need to address, and this is why we refrain from addressing this area upfront, as part of the mandatory curriculum. Nonetheless, there is a very high demand from a large portion of the student body to explore these areas. For this reason, we provide an optional module to entice part of the students in these directions. This last module is structured more along traditional academic standards than the main parts of the course which are focused on: compliance, deontology, research ethics, and the 3 major ethical challenges faced in practical settings: immaterial value, the redefinition of communication modalities and the redefinition of the private self.

## 2.2 Expectations of stakeholders

### 2.2.1 Faculty

Initial reactions of part of the doctoral council on the prescription of a mandatory course were mixed. Some professors argued, with reasonable arguments, that scientific integrity and ethical conduct are more the result of satisfying work conditions and example setting than a topic that ought to be taught, especially if this course should be made mandatory for graduation. Information Ethics as an elective seemed acceptable, but not as a large-scale course. They advocated for a simple deontology class, following strictly the model of compliance training.

Other members of the council, trained by their study area to deal in practice with complex ethical issues, were aware of the risks involved in their research area, and rightly observed that a course on the subject cannot be relevant without applied case studies. This would require a very complex deployment on the scale of a doctoral school and seems impossible to achieve with a reasonable budget. For a doctoral school of 800 students, 6 hours of face-to-face classes in small classes plus the correction of dissertations or presentations would represent several full-time teaching workloads. In addition, since doctoral students are spread among many scattered teams, organizing schedules would be an insurmountable headache.

Finally, there is an important difference between this type of course and a classic university course: in a university course, a failure is sanctioned by a bad grade, but only the student is penalized. In the case of a course on scientific integrity and research ethics, what can a lack of attention or competence mean? A priori, it is the community itself that is penalized. The course therefore provides a certification, which is mandatory but can be obtained by all, and not a 'unit of value' validating a particular competence.

Because of these tensions, the course we propose necessarily chooses a middle ground, creating dissatisfaction from all points of view - too long for some, too short for others - but will nevertheless meet all minimum expectations.

### 2.2.2 Students

Based on the outcome and feedback received in the first sessions, we can loosely classify the students who have taken the class in 4 categories:

- 60-70% of sufficiently motivated students, who did just as asked, and did in fact learn something from the course, to varying extents. About 10% of them got seriously interested in the area and are willing to learn more on the topics explored in the course. This is the middle ground we targeted, and we think we have reached our expectations for them.
- 10-15% of highly motivated students with a capacity for dialectics and verbal reasoning, sometimes having had training in ethical reasoning beforehand. Some of those felt (rightly) underwhelmed by the course format, save for a few sections.
- 10-15% of "rebel" students, who are sometimes overconfident in their understanding of Ethics and ethical reasoning. Even though some are still able to pass and acquire the course content, others cheat their way through (a few students have confessed having shared quiz results instead of studying the material). In the corporate responsibility practice, these profiles are sensitized to deontology and ethical demands through yearly repetition of short, mandatory classes. Still, it is paramount to bring forward the responsibility of the doctoral advisors of the students in this process, and this is far from acquired.
- 10-15% of the students had real difficulties with the written material and the argumentative approach, and barely passed the course. The few students who did not finish the course may be in this category too and were too shy to ask for help. For those students, the "utilitarian" approach leveraging short courses with very concrete, down-to-earth use cases and precise rules and guidelines is best suited, as is well known in the corporate responsibility/compliance practice.

Utilitarian approaches cannot be satisfied with a course that would leave aside more than 25% of its intended audience. If the goal is indeed to minimize risk exposure in a measurable way, the audience that would not be properly reached would constitute isolated areas of higher risk, rendering the whole effort moot. Therefore the non-optional part of the course takes a reductionist approach, streamlining ethical dimensions into a limited number of identifiable concepts, but is very strict that all course takers must demonstrate a minimal level of understanding. This comes at the cost of hiding the potential complexity of real-life situations and leaving them to be explored.

**2.3 Design choices**

*2.3.1 Cultural context*

The content itself focuses on the French and continental European cultural context, especially in its legal and work culture aspects. However, in information ethics, one must recognize the dominance of Anglo-Saxon works, their anchoring in concrete issues, and the relative marginality of works in French (notable French authors publish in English). This makes us privilege Anglo-Saxon authors and thinkers, such as Luciano Floridi, Lawrence Lessig, Tristan Harris or danah boyd, but also include continental thinking, with Rafael Capurro and Antonio Casilli. Finally, an attempt is made to include the point of view of emerging countries such as China or Japan, strongly influenced by Confucianism, and destined to play an important role in the decades to come. As an example of the advance of Anglo-Saxon countries in digital ethics, in the United States, the type of course we offer is integrated into the curriculum at the bachelor's degree level.

*2.3.2 Form*

The course format is strongly inspired by the courseware used in the world of compliance and corporate ethics. However, because the target audience is more homogeneous, a written format has been chosen and the level of abstraction and length of the course have been significantly increased. One advantage of written text over video is that it is easier to update in a rapidly evolving field. The different chapters move up in abstraction as the course progresses, beginning with a focus on compliance, supplemented by concrete examples, and ending with open-ended issues, refraining from definitive conclusions to encourage reflection and debate in the discussion forum.

As far as references are concerned, they are all, as far as possible, included as hyperlinks in the course (which sometimes poses problems of missing links). Whenever possible, we prefer to point to wikipedia rather than primary sources, to maximize the neutrality of the course. Thus, if a reader judges our content to be biased on a subject that refers to wikipedia, we invite the contradictor to edit the source reference itself rather than focus their criticism on the course.

The evaluation is done by closed-choice questions, which is a questionable decision, and probably the greatest difficulty in developing the course. Indeed, despite a limited number of possible choices, most questions require research and reading of documents to be answered and require analysis. Or they emphasize the need to understand the speaker's point of view to formulate an accurate answer. Several students described some questions as stressful, a sign that they do indeed provoke a wished introspection. There are about 100 questions, and half of them provide an opportunity to point out the subtleties of ethical, moral or legal issues. Student feedback confirms that many students have gained a much better understanding of the issues and the complexity of ethical analysis by trying to answer the questions. On the other hand, for those students (about 15%) who fail to reach 75% correct answers, the catch-up is very easy: we just ask them to formulate in their own words the reasoning that led to the answer that was deemed incorrect. Then, as often as possible, a discussion on the subject is held, with several students participating. All participants receive a maximum score on this question regardless of their final opinion. This is what allows us to introduce our attendance to ethical analysis, in a process where the discussion is animated by students, rather than directed.

*2.3.3 Towards a co-construction of values and ethical practices*

Measuring the effectiveness of the course is a fundamental issue in corporate compliance and ethics. As this field has more than 20 years of history, it can begin to benefit from final indicators on the evolution of observed infringement rates, the effective implementation of the rules taught, or traces of discussions and referrals to ethics committees. Since this course is new, we have limit ourselves to primary indicators: success rate of students, estimated fraud rate (around 5-10%), student satisfaction and feedback, number of direct dialogues generated, and finally, number of students enthusiastic about the course and willing to invest in the study and teaching of these subjects (~5%).

This last point is important because it prefigures the long-term future of this subject area at the university. Indeed, the course as designed remains subject to the biases of its designer and reviewers. It is also unlikely to evolve fast enough to take into account the fast evolution of its main subject. This is why the long-term objective is to bring it to an open platform, like wikipedia, and to entrust it to doctoral students and researchers so that it constitutes a living and co-constructed information base on the subject of scientific integrity and digital ethics. It is by transforming learners into actors that we can be sure that teaching will bear fruit and be effective.

We can believe in the possibility of success of this enterprise by an interesting point to note: this course being mandatory, it constitutes a meeting point for doctoral students, an opportunity that is too rare otherwise. Thus, we have seen the development of exchanges between doctoral students on related themes but working in different laboratories, which we hope to see develop. Finally, the observation of drop-outs - about 5% while certification is within everyone's reach - makes it possible to identify difficult thesis paths, pointing the doctoral school's managers on the opportunity to accompany certain isolated or disoriented doctoral students.

Thus the title of this document, which aims to replace a 'teacher/learner' vision with a co-constructed space for sharing common ethical values and practices, is justified. However, this objective still requires relatively substantial investments, which the doctoral school may not be able to provide. It is not difficult to broaden the audience of the course with the current means of sharing, which would make it possible to find a way of financing this undertaking.

## 3 A COURSE PROPOSITION

The course starts with a pragmatic "must know" perspective, which can and must be acquired easily and fast, and progresses towards more conceptual developments aimed at opening up to the uncertainties and risks that are specific to the student's scientific domain:
- The Doctoral Contract (researcher as a subject under law and an employee),
- Research Integrity (producing science correctly),
- Research Ethics (producing science responsibly),

- Computer and Information Ethics (how these apply to your research context),
- Intellectual property (researcher as a producer and consumer of value),
- Scientific Communication and Internet Ethics (disseminating knowledge publicly),
- Privacy and Personal data (protecting our digital selves),
- Emerging issues in Computer and Information Ethics (what's next, optional chapter).

Each of these chapters includes:
- A lesson to be read by the student. We trust our target audience to read and learn from a text. As much as possible, the content is summarized in a set of checklists involving as few guidelines as possible, for easy memorization.
- Examples of how the material discussed may be relevant in their context of work. We provide "horror stories" and "happy stories" of past events relevant to the topic addressed, as well as questions and issues for the student to ponder over.
- References are hyperlinked in the course of the text, and a "further reading" section is sometimes provided. When Wikipedia provides adequate content, we favor using it rather than more volatile, single language and possibly less consensual sources.
- Finally, some validation questions are proposed after each section. To allow self-delivery of the material, the exercises are proposed as multiple-choice questionnaires, which enable full automation of the test administration and certification process. We are careful however that the questions cannot be answered properly without students undertaking their own investigations and analysis of the material provided, to acquire some practice in the manipulation of the tools and concepts we provide.

A discussion forum is provided, where students can ask questions, present their opinions on various topics, and finally obtain credit for the questions they miss, as the goal remains to certify all course-takers.

The introduction is intended to make doctoral students aware of the purpose of the course. A priori many of them find themselves in the situation, new for a student, of being in an activity of *production (of knowledge and/or value)*. This context creates new responsibilities towards colleagues, employers, and even science and society in general. It is important for the doctoral school to make sure that they are aware of these new demands on them, as a lack of knowledge can have harmful effects. The purpose of the course and the reason why it is mandatory is therefore the protection of the doctoral student, his or her entourage and society in general. A few striking examples of blunders or inappropriate behaviors that have had a big impact are briefly presented (Alexander Kogan delivering his research to Cambridge Analytica and the "Morris Worm").

### 3.1 The doctoral contract

The first chapter focuses on compliance, which we believe is a prerequisite to any form of ethical training that targets a wide audience. It reviews some of the foundations of law, how it is created and how to get informed advice; the main provisions of the employment contract (principle of subordination, intellectual property, time management…), various aspects of internal regulations and the life of the laboratory, as well as the main legal acts that may intervene in the course of the work (confidentiality agreements, multiparty agreements…), and their effective consequences. This is to ensure a minimal common base on the subject.

At the suggestion of students from the first session, we have added a section on conflict and harassment management, as well as a section on psychological support (remember that doctoral students are among the categories of population most prone to psychological suffering): as mentioned above, a necessary precondition for maintaining scientific integrity and research ethics is simply well-being at work and in one's life. It is hoped that this section can be a modest help for doctoral students who may face hardship.

The evaluation is made through tests of reading contracts and searching for legal information. Even if this chapter does not 'fly high', it is greatly appreciated by some students, especially foreign students, because it provides information that is too often considered obvious, even though it is not always known.

### 3.2 Scientific Integrity: Producing Knowledge Correctly

Scientific integrity is first about understanding the basics of the scientific method, and the rudiments of epistemology. This section, therefore, in a Popperian perspective, reminds of the goal of science as knowledge creation through a proper process. It reminds the necessity of distinguishing knowledge creation from transformative intents, which are often at the origin of a scientific enterprise, particularly in health and engineering sciences. This distinction plays an important role in research ethics, and it is important to bring it forward early enough to raise it when necessary. Next, a short reminder on the scientific method is presented, which imposes rigor and reproducibility. This allows introducing work practices requirements, including being attentive to biases and methodological errors. Then, a discussion on the production and dissemination of results, allows introducing the necessity of creating a distance between the objective results and their interpretation within and outside the scientific context.

The second part of the chapter presents and explains the different kinds of improper conduct, voluntary or not, that the researcher is likely to encounter in his or her career: plagiarism, falsification, sabotage, etc. Above all, it provides clues to identify and prevent them while respecting ethical conduct. Finally we provide a checklist, in the hope its memorization can help students maintain a proper deontology:

- respect the goal of science (knowledge acquisition vs transformative intent)
- respect the methods of science (validation/refutability), notably reproducibility
- respect fellow researchers & apparatus of science (cite, prevent plagiarism, provide context to let one assess bias exposure)
- be responsible: validity of results, claims and acceptance of limitations
- responsible disclosure, avoiding mistakes and bias.

The assessment is a test of basic knowledge in epistemology and a presentation of cases of fraud and types of bias that the student must identify and classify.

### 3.3 Research ethics: producing knowledge responsibly

This chapter is the longest and most ambitious part of the course. We provide a general overview of the issues at stake in research ethics, rooted in the history and practices of different scientific fields, it attempts to subsume these issues into a synthetic and readable set of principles applicable in any context. It is a perilous exercise to allow oneself to simplify this complex subject. However, the definition of a space for reflection with a limited number of dimensions is greatly appreciated by students and is particularly structuring. Research ethics can be summarized with the Hippocratic tenet: 'First, do no harm'. This maxim comes from medical ethics and one cannot consider this chapter without giving it a central role, mentioning, among others, that the terms "bioethics" and "research ethics" are used synonymously by many specialists, and even by search engines!

The first section thus approaches bioethics from a historical perspective: Hippocrates, the Nuremberg Code, then subsumes the latter as three first major directives on what "First, do no harm" means:
1. Respect for the human being, including ensuring the free and informed consent of stakeholders.
2. "Due Diligence": devote your knowledge and energy to the protection of the subjects first and the progress of the project second. Do not change priorities without involving stakeholders.
3. "Precautionary Principle": maintain a permanent assessment of the risk/expected benefit ratio and use this analysis in decision making.

A fourth guideline, stemming from the Helsinki Declaration and other references is proposed: the principle of 4. openness/transparency, which requires scientists to explain the motivations of their projects and contextualize them. Finally, this section concludes with a brief presentation of the mechanisms in place, such as ethics committees, to ensure that these principles are respected, balance points of view and resolve ethical tensions.

The second section goes through other scientific fields to identify ethical issues at stake, assimilating them to the dimensions previously seen or proposing new ones. Research using animals and oil research allow us to generalize the first dimension: it is about respecting the subject of study, with its own characteristics, and not only humans: minimizing suffering in the case of animals, minimizing environmental impact for oil research. Examining social sciences, and in particular economics, allows us to identify another subject: economic and social structures. This leads us to identify again the need to distinguish the epistemic role of science from transformative intents, such as the interpretation of scientific results for policy making. This highlights the need to establish lines of responsibility to distinguish roles, in what is called the principle of 5. independence. Finally, we show that our picture is not complete, mentioning among other things an issue that is not covered, such as the need to 6. respect social norms, as shown, for example, by the delicate question of stem cell research.

As in the first chapter, for memorization and to allow structured investigation process, we propose a checklist, or "reading grid" to identify ethical dimensions:
- Respect for the subjects:
    o Humans
    o Social – economic structures
    o Sensible beings & Nature
    o (Informational: addressed in next chapter)
- Due Diligence
- Precautionary Principle
- Transparency
- Independence
- Respect of social norms

The third section is an introduction to ethical analysis, mentioning the traditional classification of ethical systems (virtues, teleological, deontological, but also Confucianism…), then presenting how to build an analysis, integrate all stakeholders to lead to consensual and appropriate decisions, and also accept the constraints that this approach poses.

The validation focuses on an interpretation of various ethical and deontological codes following this reading grid (codes of ethics of New Zealand anthropologists and the ACM). At this stage, we are not yet able to put ethical analysis in context, as we have not yet tackled any topics that would challenge the students' expertise to raise ethical questions about their own field of study.

### 3.4 Information ethics

Equipped with a landscape of research ethics and a framework, we can now focus on ethical issues specific to the field of study of students in Information Sciences and Technologies. To do so, we draw up a panorama of the properties of computer systems: perennial/transient, local/global, explicit/implicit data, deterministic/non-deterministic, autonomous/controlled... These properties distinguish them from the physical world, the social world and the world of ideas, thus justifying the consideration of a new, informational space. Information ethics, as proposed by Luciano Floridi, aims at preserving and improving the well-being of the *infosphere*.

To organize a proper ethical reflection space, we are interested in the interactions between the infosphere and the material and human world, identifying domains of risk exposure. Each of these domains leads to one or more chapters in the follow up of the course:

#### 3.4.1 The object of study: computer systems, data & networks

Of course, research cannot compromise the security and integrity of computer systems in production, which requires a section to ensure that students adhere to appropriate security standards. This section also introduces codes of conduct for cybersecurity research, authorizing, under certain conditions, intrusion tests, in application of the principle of independence (white hat/grey hat).

#### 3.4.2 Society and economic players

Information technology has an undeniable economic value. It is up to the researcher to protect this value, which will lead us to develop the chapter on intellectual property. In the same way, it transforms our societies, opening the need for a reflection on its impact, in what is called "internet ethics", developed in a chapter where the deontology of scientific communication is presented, as an introduction to the subject.

#### 3.4.3 The human

Finally, digital technologies are redefining the limits of the self and relationships between people. How do we protect our digital identity? This is the subject of the chapter on the protection of personal data and privacy. The chapter outlines the DGPR as conceptual framing and extends these with short presentation on unresolved dilemma in this space.

#### 3.4.4 On the transformative nature of information technologies

Beyond these, digital technologies create profound transformations. In the last chapter of this course, we explore a number of current directions in information ethics, subjects of important work and reflection, but on which few precise recommendations can be made, except that the researcher bears responsibility for their development. It is therefore his duty to keep aware of and inform society about the changes they bring about, in terms of automating decisions (algorithm governance), capturing attention (design ethics), data and algorithm governance, mastering autonomous systems, and finally, containing the digital divide. Each of these directions is presented briefly, with a few pointers for further reflection, in the last chapter, which is optional.

The evaluation of this chapter is a continuation of the chapter on research ethics, specializing in questions on digital ethics issues, this time introducing real case studies.

**3.5  Intellectual property: protection of the value produced**

The value produced (and consumed) by science in general, and by digital technologies in particular, is to a very large extent immaterial, be it models, data, software or even prototypes. This value is therefore protected by the intellectual property regime, the basics of which it is important for doctoral students to master, both to respect the value produced by others, and to secure the value they produce.

The first section introduces the basics of the different intellectual property regimes, essentially copyright and patent law, which are those most likely to be useful to researchers, with the following outline:
- Intellectual property regimes
- Copyright/copyright; exceptions to copyright. Main recommendations for the protection and dissemination of its productions, respect for the intellectual property of others.
- Patents: fundamental principles, including formulation of claims. Prior art search, short introduction to patent filing.
- Valuation of patents: different economic models in use.

The second section is an essay on various issues raised by the Internet and software to the bicentennial principles of intellectual property: the place of free software, the controversial aspects of software patents, the different European and American approaches to the dissemination of works on the Internet, in particular the role of platforms, the disruption of related business models, the puzzling issues raised by AI-generated content...

Validation is based on classical questions of copyright exceptions regime, interpretation of patent claims...

**3.6  Scientific and ethical communication on the Internet: protection and limits of free expression**

As introductory matter, this chapter explores the relationships between the doctoral student, his/her team, his/her research community, and the outside world:
- Presenting results and submitting an article
- Participating in research evaluation
- Presenting work and addressing the public (in person or online, for example on a personal website, blog or forum).

These bases of deontology are used in the second part to explore the complex issues of Internet ethics, the delicate balance between freedom of expression and respect of others, particularly for Internet platforms and search engines (search ethics), at a time of important debates on online manipulation of opinion. The points of view are supported by essays and internal documents from platforms presenting the recent evolution of their policies on these subjects.

The evaluation focuses on the deontology part, reading tests of internet platforms' 'terms of service', and discusses an interpretation of complex ethical issues, such as the policy of content censorship on google and facebook platforms.

### 3.7 Personal data and privacy online: protecting our digital identities

This chapter presents the protection of the digital self, first from a legal and deontological point of view, with a presentation of the rights to be protected, and of the RGPD, encouraging doctoral students in particular to get involved in this promising field of digital ethics (on the proposal of the CNIL, the French DPA).

The framework is then broadened to present the complexity of the ethical issues at stake, the tensions between the technological world, where all data are equal, and positive law, where this is not the case, and the exceptions authorized for research purposes, notably in application of the principle of research independence, described above.
Finally, we conclude with a presentation of the tools and approaches to personal data protection that can and must be implemented by the researcher using this type of information.
The assessment focuses on a few cases of interpretation of the DGPR, as well as clear examples of potential violations of privacy, which may or may not fall within the scope of this regulation.

### 3.8 Emerging Issues (Open Areas of Digital Ethics)

Up to here, the course, even when dealing with open questions or unresolved ethical tensions, has always offered a concrete set of recommendations and methods to address dilemmas that may arise. The last part of the course, which is optional, presents more open-ended issues, on which there may be recommendations or a legal framework, but no synthetic approach or recommendations that we would be able to make explicit.

This chapter proposes 5 essays on current themes in information ethics:
- Design ethics, digital addiction and opinion manipulation.
- Algorithm governance - decision automation
- Ethics of Data Science and Machine Learning     ⎤ AI Ethics mixes concerns of those 3 areas.
- Ethics of autonomous systems (& robotics)      ⎦
- The digital divide and intercultural information ethics

Approximately 40% of the students take this part of the training and do the evaluation (which is easy), although it is well indicated that it is optional.

### 3.9 Conclusion of the course

To conclude the course, we propose a summary of the main lessons learned, in the form of 6 main recommendations and a participatory approach to identify and address ethical issues.
*A particularity of information ethics is that solutions to the problems posed by new technologies are, at least in part, technological in nature. While society tends to regulate its ethical dilemmas through law and morality, digital technologies are as much part of the solution space as they are part of the problem space. Much like a padlock is a technological means of enforcing property ownership and privacy.*

# 4 RECEPTION OF THE COURSE

## 4.1 Feedback from the students

The online course has now been taken by about 2000 students, most of them as part of a mandatory curriculum in scientific integrity and research ethics. In the first few sessions, we gathered detailed feedback, which we present now. The following analysis was made at the end of the first session, in March 2018 (70 answers out of 98 course takers). Since then, we have greatly taken their feedback and improved the course, but the overall trends are maintained.

### 4.1.1 Course appreciation

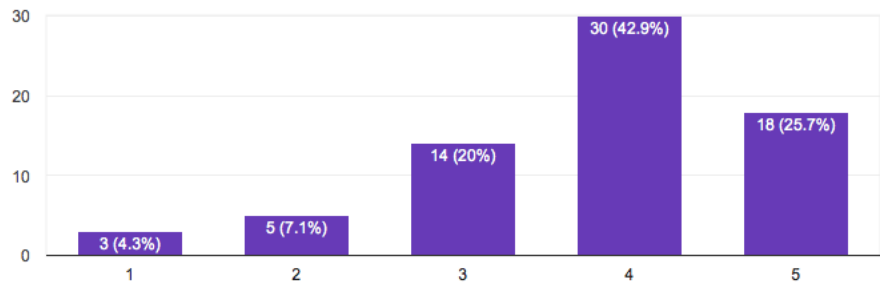

There is a clear consensus that the material taught must indeed be known, save for 11% who may consider ethical reflection as more of a luxury than a need in their research work. This view is confirmed by the next question:

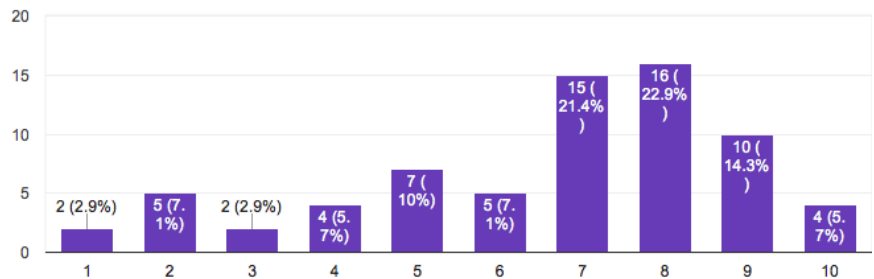

The 13% who consider they have not learned much overlap significantly with those who consider the topics presented are useless to start with. Some also don't like the fact that Ethical reasoning has been approached via guidelines and quizzes that require interpretation rather than via a verbalized, dialectical approach.

Approximately 23% of students give an average rating. From informal discussions, we infer that either they had an awareness of Ethical subjects and found the course oversimplified, not meeting their expectations, or they acquired the course passively and genuinely show little concerns for the topics taught. For instance, their thesis area is very theoretical, and they don't feel like their work could ever present deontological or ethical risks. Finally, some students may be frustrated by the lack of more modern approaches to teaching, involving more interactivity. As one of the students puts it: "a passing grade can be obtained with just a careful reading of the material, without exercising our real understanding. This is not interesting." (Author's note: plenty of other students have mentioned that the course was very hard, and even stressful!).

Almost 2/3 of the students give a rating of 7 or more, which lets us consider the goal of reaching the "silent majority" with effective material is accomplished.

The correlation between perceived usefulness for everyone (Global importance) vs. personal usefulness is somewhat clear, and students who finished the course later appear to have higher appreciations (darker, smaller circles are older entries in the feedback form, wider/lighter entries are more recent):

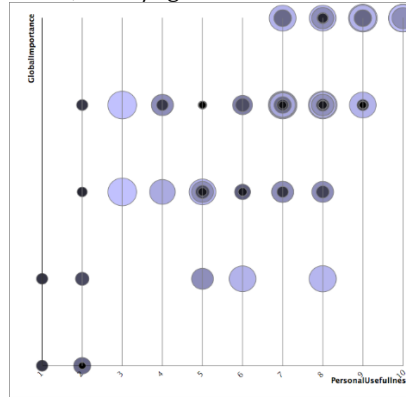

### 4.1.2 Course content

Your preferred chapter or section (number of mentions)

Chapters:

| | |
|---|---|
| Doctoral Contract | 5 |
| Research Integrity | 11 |
| Research Ethics | 22 |
| Computer and Information Ethics | 7 |
| Intellectual Property | 11 |
| Scientific comm. & internet ethics | 7 |
| Privacy and Personal Data | 11 |
| Emerging Issues | 11 |

Sections:

| | |
|---|---|
| Introduction | 1 |
| AI & Robotics | 2 |
| Philosophy | 2 |
| Trolley problem | 2 |
| Bias & Stats (integrity) | 1 |
| Threats to information systems | 1 |

The lesson to retain is that students have varied tastes and were definitely not looking for the same things in the course.

Chapters/section that could be removed (number of mentions)

| | |
|---|---|
| Doctoral Contract | 8 |
| Research Integrity | 3 |
| Research Ethics | 3 |
| Computer and Information Ethics | 2 |
| Intellectual Property | 3 |
| Scientific comm. & internet ethics | 3 |
| Privacy and Personal Data | 3 |
| Emerging Issues | 1 |
| Machine Learning | 1 |
| Robotics | 1 |
| Philosophy & Trolley problem | 2 |
| Reduce overall size | 2 |

Interestingly, except the first section, which has more of a compliance bent, the results are spread and not much information can be drawn from the poll: once again, students have varying opinions of what matters most in an integrity and ethics course. Of course, the first section is an integral part of what must be known by any professional (not just researchers): it is the minimal platform on which "immediate utilitarian/deontologists" will settle, and thus it is here to stay. The students who want to get rid of the central chapter, on Research Ethics per se, tend to be those who dislike the subjective nature of the area, and resist the notion that ethical questions have no clear-cut answers and require interpretation. Sadly too, the chapter that the students want to get rid of the least, "Emerging issues...", is the one that is optional!

Someone mentions: "All sections which are too much subjective, by turning questions in a ways that we can find the information somewhere clearly". This indicates someone for which an approach based on simple rules and guidelines is mandated rather than an introduction to ethical analysis.

### 4.1.3 Didactics

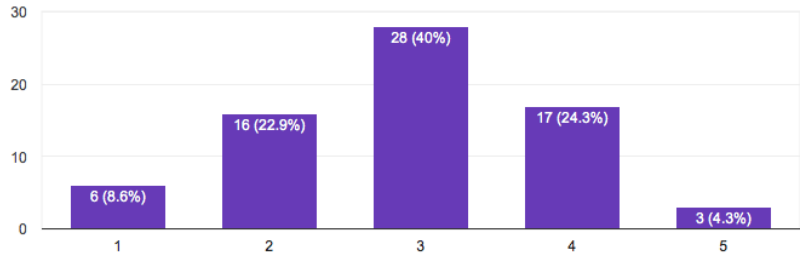

Once again, somewhat a Gaussian, but not as positive a result as the previous one. Yes, we'd like to do better, in particular to prevent the possibility of cheating, but this will take time to perfect.

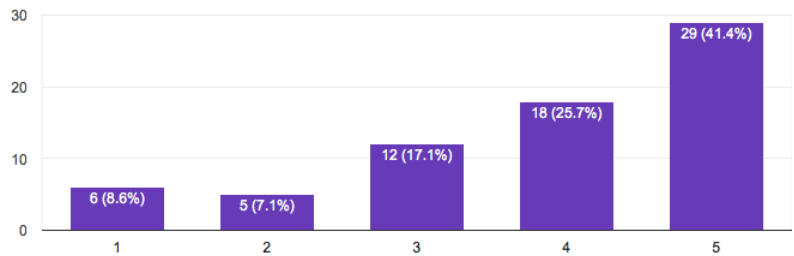

Given the constraints presented in the introduction, and the overall response to this question, this format is here to stay. Definitely, more interactivity would help many and allow to explore Ethical analysis more thoroughly. But this would be much more difficult to setup and demand that students write essays, which I suspect many don't want to do.

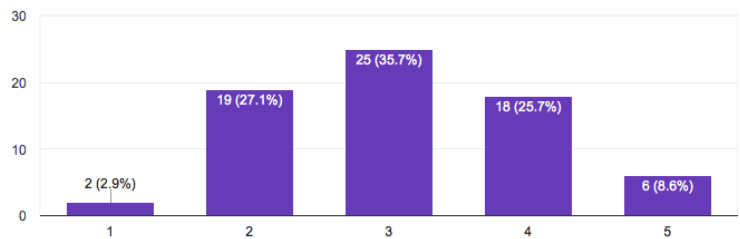

*4.1.4 Questions difficulty assessment*

We have examined the success rates of the 96 validation questions, which range from 22% to 100%. Questions whose success rate is below 50% (8 questions) will have the number of allowed trials increased or will be carefully reworded. Questions which have a success rate above 90% will see their difficulty slightly increased on a per-need basis to compensate. We will make a best effort to remove the "other" option in the subjective questions, or provide better hints when this "other" option should really be considered.

## 4.2 Feedback from the teaching body

The course was initially proposed as an alternative to a more classical, less intensive course (one afternoon conference) on scientific integrity and research ethics, that did not focus on information ethics. Students were free to choose between either classes. Despite its increased difficulty and demands, our course received much better appreciation, for a large part because it focuses on the specific issues and uncertainties that pertain to the scientific domain of students. Each session, several researchers take the course for their own education, and the curriculum and its design have now been presented in multiple seminars and conferences, and there is an increased demand for expanding the audience of the course to other contexts, in various universities and engineering schools in France.

## 5 CONCLUSION

## 5.1 AI Ethics and its place in a general curriculum on Technology Ethics for STEM disciplines

It may have been noticed by now that our curriculum merely brushes the subject of AI Ethics, in spite of the large media coverage of the topic and the abundance of initiatives aimed at progressing in this area.

There are 3 reasons for this:
1. the goal of the course and expectations of the audience do not call for it,
2. maturity of the area of AI Ethics, and finally,
3. a design choice to structure the course according to the human domains of activity that are affected by the evolution of technology, rather than along a techno-centric perspective, listing technologies and the potential issues they create.

*5.1.1 Course objectives & AI Ethics*

As must be clear by now, our objective is not so much to provide technical tools to our students as it is to
a) introduce a fairly rigid compliance framework that can serve as a smaller common denominator for dealing with ethical issues,
b) sensitize them to the responsibilities their work entails
c) introduce them to dialectical reasoning and dilemma resolution, as well as constructing collectively a set of shared values and practices.

While AI Ethics is certainly an important area to investigate, at this point the guidelines and common practices are too elusive to provide such a conceptual framework grounded in practical use cases. Few students have had the opportunity to deal *in practice* with delicate issues such as algorithmic biases and fairness. Also, unlike the core topics we address in the course, not all students will have an opportunity in their career to address typical AI Ethics dilemma. This justifies relegating this area in the optional part of the course, for the students whose work domain involves issues related to AI, or who consider following a specialization in Information Ethics.

### 5.1.2 Maturity the area of AI Ethics

The area of AI Ethics lacks a clear delimitation, that could be epistemologically grounded in immutable concepts. AI now has little to do with what it was 30 years ago and is vastly different again from what it was at its inception 60 years ago. 30 years ago, books on AI talked about inference systems, language grammars, symbolic pattern matching and possibly ontology development. 30 years ago, computer vision, voice recognition and even natural language processing were considered 'side quests', only distantly related to AI, and built on top of the 'signal processing' academic community. "AI" meant the imitation of reason, not the imitation of perception. If we want to provide firm concepts to young professionals, we need a clearer delineation of the domain of AI and by extension, of AI Ethics. While the issues dealt with by expert groups such as the EU HLEG on AI are very real, most of the visible media activity tends to be sensonialistic, drawing the attention on caricatural interpretations of actual dilemma, that are far remote from the true concerns of engineers and scientists. A typical example is the naïve presentation of the trolley dilemma applied to the autonomous car, neglecting the fact the engineers' goal is not to chose between deontology or teleology, but rather to never put the car in a situation where it may have to choose, which leads to a host of other issues, such as the possibly insurmountable difficulty of inserting a car seamlessly in human-dominated traffic.

For a graduate level course, we cannot address such issues with similar artificial examples, which would lead to a false impression of the real challenges and of the effective methods to tackle them.

### 5.1.3 Socio-centric vs techno-centric approaches to information ethics

As mentioned earlier, most of the ethical dilemma we face do not concern a particular technology, but rather how it is to be put to use. The very same algorithm can be used to recommend a movie on a streaming website or to recommend a prison sentence, and yet, the ethical implications will be vastly different, and will be resolved with vastly different approaches. Conversely, many of the ethical concerns involved in letting an algorithm sentence a prisoner are indifferent to the type of algorithm being used, be it statistical or procedural. We once again state that, in our opinion, ethical methods and tools must be presented and organized along socio-centric lines rather than techno-centric lines.

As a consequence, instead of proposing a specific section on this subject, we distribute the topics usually discussed in AI Ethics in 3 distinct chapters, that do not specifically focus on AI technologies: Algorithm governance and decision automation, Data science and machine learning ethics, and the ethics of autonomous systems. Then, a small section introduces the main ongoing initiatives in defining guidelines and regulation around the various subjects addressed under the umbrella of AI Ethics.

## 5.2 Future directions

The course is now more or less stabilized in its MOOC form, and is scaling up, with the latest session accepting over 900 course takers from France and the rest of the world (though, for a large part, this is because the course is free, and only 342 participated effectively). Graduate education in scientific integrity, research ethics and information ethics can probably be performed with a similar course design.

Several components are however missing to provide what in our view would be a proper educational system for this purpose:

- A resource center, possibly inspired from (or piggy-backed on) the ACM resource center on computer ethics [1]. This resource center should provide a platform similar in essence to Wikipedia to enable the collaborative construction of shared values and practices to address the ethical challenges of the digital revolution.
- Discussion forums enabling transparent exchange on the various topics of information ethics that need to be settled. We are attempting to create such a forum on https://reddit.com/r/ComputerEthics but so far participation is weak.
- A certification authority and a smaller online MOOC that would provide a deontology-oriented, base certification, which would be made mandatory for graduate level researchers. Inspired from compliance education, this certification would be extended yearly with short "booster" trainings of one hour at most, each focused on a case analysis and its resolution.
- The present MOOC should be enhanced, provide more interactivity and train test takers to ethical case analysis. Its aim would be similar to the CertNexus [5] course, delivering a certificate of "Ethical Technology Specialist".

If this works, this type of thematic site could prefigure the "ethics committee 2.0" of the future, using current technologies to monitor and raise awareness, enriching the debate with multiple voices, while proposing open arbitration and supervision mechanisms, a central function of the ethics committees.

**ACKNOWLEDGMENTS**

The author wishes to thank the STIC doctoral college of University Paris-Saclay for having entrusted us with such a responsibility, as well as the numerous contributors and course takers who have provided much insightful feedback to continuously improve the course and extend its audience: we thank Christine Froidevaux and Arnaud Billion for contributing some of the course material, the reviewers of the course: Alice René, responsible for bioethics regulation at CNRS, Max Dauchet, professor emeritus at the Université de Lille and former president of the CERNA, Jules-Marc Baudel, former member-at-large of the Paris bar association, Laurence Devillers, Research Director at LIMSI and member of the CERNA, Wendy Mackay, Research Director at INRIA, Michel Beaudouin-Lafon, professor at University of Paris-Sud, Sophie Vuillet-Tavernier, director of public and research relations at CNIL, and Russell Owen, Engineering Manager at Lyft, for their numerous and insightful suggestions. We also thank the student beta-testers Steven De Oliveira and Maxime Martelli, as well as the teaching assistants: Aylen Ricca, Cécile Moulin, Philip Tchernavskij and Ludovic David. Finally, Nicole Bidoit, Hanna Klaudel and Alain Denise, at the head of the University of Paris-Saclay STIC doctoral college, have strongly supported this undertaking. Finally we thank Rafael Capurro for his insightful comments on the drafts of this paper.